\documentclass[%
 reprint,
superscriptaddress,
noeprint,
nodoi,
 amsmath,amssymb,
 aps,
]{revtex4-2}

\usepackage{graphicx}
\usepackage{dcolumn}
\usepackage{bm}
\usepackage{xcolor}
\usepackage{siunitx}
\usepackage{upgreek}
\usepackage{comment}
\usepackage{xr}
\usepackage{hyperref}
\usepackage{pgf}
\usepackage{tikz}
\usetikzlibrary{arrows,automata}
\usepackage[latin1]{inputenc}
\usepackage{placeins}

\usepackage{mathtools}
\DeclarePairedDelimiter\abs{\lvert}{\rvert}%
\DeclarePairedDelimiter\norm{\lVert}{\rVert}%

\makeatletter
\let\oldabs\abs
\def\abs{\@ifstar{\oldabs}{\oldabs*}}
\let\oldnorm\norm
\def\norm{\@ifstar{\oldnorm}{\oldnorm*}}
\makeatother
\begin{document}

\preprint{APS/123-QED}

\title{Squeezing and multimode entanglement of surface acoustic wave phonons}

\author{Gustav Andersson}
\altaffiliation{Present address: Pritzker School of Molecular Engineering, University of Chicago, Chicago IL 60637, USA}
\email{gandersson@uchicago.edu}
\thanks{These authors contributed equally.}
\affiliation{%
Department of Microtechnology and Nanoscience MC2, Chalmers University of Technology, SE-41296 G\"oteborg, Sweden
}

\author{Shan W. Jolin}
\altaffiliation{Present address: IQM Finland Oy, FI-021 50 Espoo, Finland}
\thanks{These authors contributed equally.}
\affiliation{%
 Nanostructure Physics, KTH Royal Institute of Technology, SE-10691 Stockholm,
Sweden
}%

\author{Marco Scigliuzzo}
\affiliation{%
Department of Microtechnology and Nanoscience MC2, Chalmers University of Technology, SE-41296 G\"oteborg, Sweden
}
\author{Riccardo Borgani}
\affiliation{%
 Nanostructure Physics, KTH Royal Institute of Technology, SE-10691 Stockholm,
Sweden
}%
\author{Mats O. Thol\'en}
\affiliation{%
 Nanostructure Physics, KTH Royal Institute of Technology, SE-10691 Stockholm,
Sweden
}%
\affiliation{%
 Intermodulation Products AB, SE-82393 Segersta,
Sweden
}%
\author{J. C. Rivera Hern\'{a}ndez}
\affiliation{%
 Nanostructure Physics, KTH Royal Institute of Technology, SE-10691 Stockholm,
Sweden
}%
\author{Vitaly Shumeiko}
\affiliation{%
Department of Microtechnology and Nanoscience MC2, Chalmers University of Technology, SE-41296 G\"oteborg, Sweden
}
\author{David B. Haviland}
\affiliation{%
Nanostructure Physics, KTH Royal Institute of Technology, SE-10691 Stockholm,
Sweden
}%
\author{Per Delsing}
\affiliation{%
Department of Microtechnology and Nanoscience MC2, Chalmers University of Technology, SE-41296 G\"oteborg, Sweden
}

\date{\today}

\begin{abstract}
Exploiting multiple modes in a quantum acoustic device could enable applications in quantum information in a hardware-efficient setup, including quantum simulation in a synthetic dimension and continuous-variable quantum computing with cluster states. We develop a multimode surface acoustic wave (SAW) resonator with  a superconducting quantum interference device (SQUID) integrated in one of the Bragg reflectors. The interaction with the SQUID-shunted mirror gives rise to coupling between the more than 20 accessible resonator modes. We exploit this coupling to demonstrate two-mode squeezing of SAW phonons, as well as four-mode multipartite entanglement. Our results open avenues for continuous-variable quantum computing in a compact hybrid quantum system.
\end{abstract}

\maketitle

\section{Introduction}
Quantum computation and simulation show potential for tackling difficult computational problems by leveraging superposition and entanglement in engineered quantum devices. Although individual quantum systems can be controlled with excellent precision, scaling the hardware to the complexity required while maintaining sufficient control remains a challenging problem \cite{Krinner2019}. Most architectures proposed for quantum simulation and computation \cite{Devoret2013, Gambetta2017, Wendin2017, Puri2017,Kounalakis2018} use one circuit component for each node in the processor, leading to demanding hardware requirements for practical applications. It is therefore attractive to explore alternative approaches to quantum computing that provide for compact encoding and processing of quantum information.
\begin{figure*}[htb]
	\centering
	\includegraphics[scale=0.33]{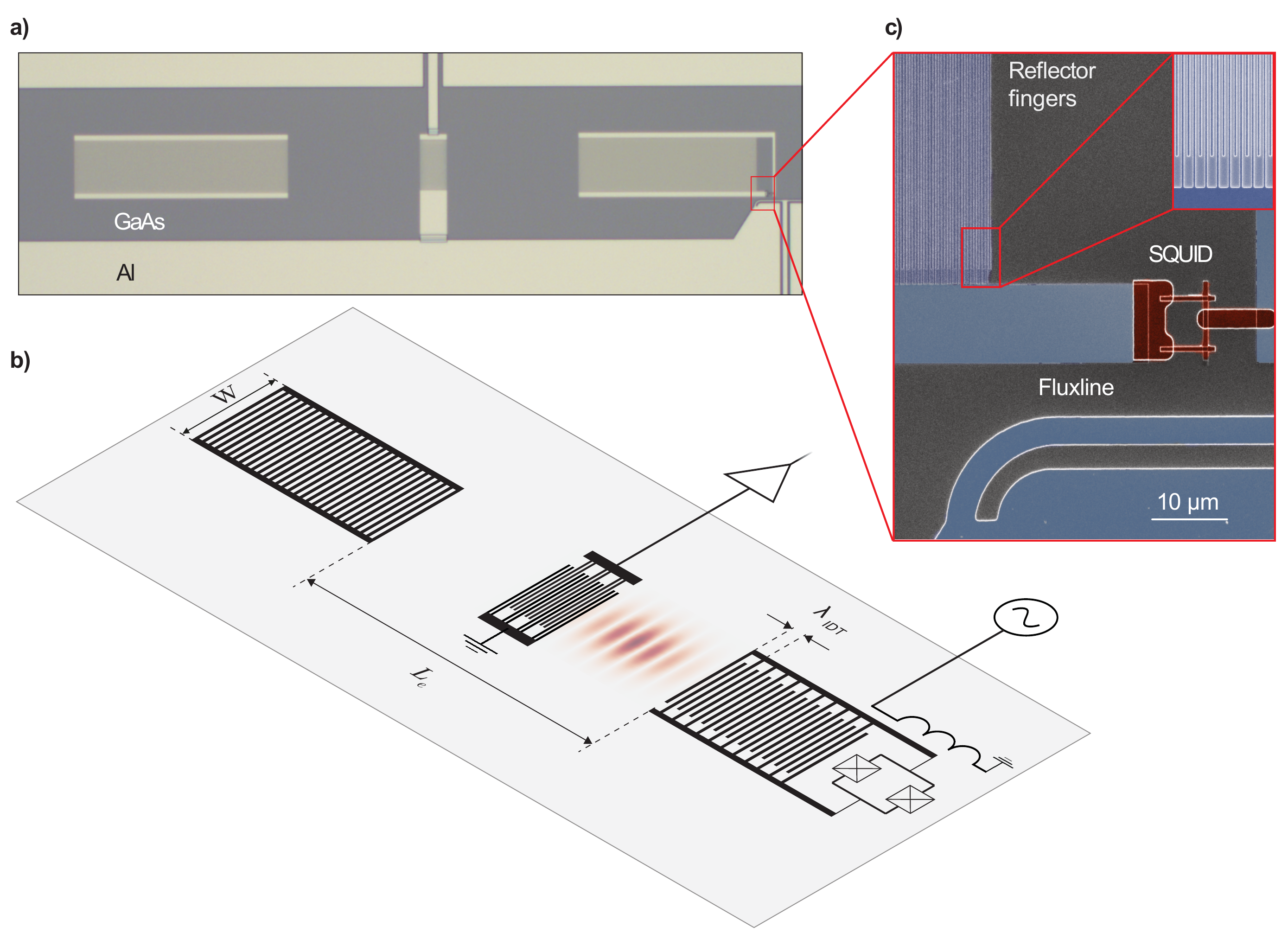}
	\caption{\textbf{Device layout.} \textbf{a)} Optical microscope image of the hybrid SAW resonator. The left Bragg reflector has 1200 fingers all shorted together. The reflector on the right hand side has 500 unit cells of fingers alternatingly connected to either the top or bottom electrode, similarly to an IDT. As shown schematically in \textbf{b)}, the electrodes are shunted by a SQUID which is modulated by a fluxline. The IDT at the center provides a single input and output port to the resonator. The unit cell of the IDT has a period of $\lambda_\mathrm{IDT} =\SI{736}{nm}$ and a double finger structure to suppress mechanical reflections. \textbf{c)} False-color scanning electron micrograph of the right hand side reflector with the SQUID and on-chip fluxline.}
	\label{fig: schematic}
\end{figure*}

In principle, the use of continuous variables (CV) allows for realizations of measurement-based quantum computing with frequency combs, requiring only a small number of coupled quantum systems \cite{Weedbrook2012,Pfister2019}. This paradigm of quantum computing relies on entangling a large number of modes rather than qubits, and does not face fundamental restrictions preventing universality and fault-tolerance \cite{Flammia2009}. Much experimental progress in CV encoding of quantum information has been achieved in the domain of quantum optics \cite{Chen2014, Araujo2014, Yoshikawa2016, Larsen2019, Asavanant2019}, where optical parametric oscillators can be used to generate large cluster states, a type of multipartite entangled states providing the resource for CV quantum computation. With superconducting circuits, CV encoding has been pursued mainly for error-correction schemes on logical qubits encoded in many-photon superconducting cavity states \cite{Vlastakis2013, Rosenblum2018}. While parametric devices are important for low-noise amplification \cite{Roy2016}, multimode measurement-based schemes for quantum computing at microwave frequencies have received relatively little attention \cite{Sivak2020, Lahteenmaki2016, Chang2018}. The dominant approach to quantum computation with superconducting quantum circuits has been the gate-based quantum processor, with most effort expended on scaling up the number of physical qubits \cite{Arute2019}.

A limiting factor for realizing CV encoding in circuit quantum electrodynamic (QED) systems is the typically large electromagnetic mode spacing, making devices with a large number of accessible modes very long or difficult to design. On the other hand, microwave frequencies are amenable to digital signal processing and thereby a greater degree of programmable control than is currently possible in optical systems. The prospect of integrating superconducting qubits as a means of providing non-Gaussian operations necessary for quantum advantage in computation \cite{Mari2012} is an additional strength of microwave circuits.

We demonstrate an approach towards realizing CV quantum computation based on cluster state generation and control in a multi-mode hybrid superconducting quantum acoustic device. The interaction between mechanical oscillators and superconducting circuits has been used to show quantum effects \cite{O'Connell2010}, including entanglement \cite{OckeloenKorppi2018}. Surface acoustic wave (SAW) resonators support dense mode spectra with high Q-factors ($ > 10^5$) and have been used in multi-mode experiments in the quantum regime \cite{Moores2018,Sletten2019,Andersson2021}. Substantial progress has also been made in recent years in the controlled generation of non-classical phononic states \cite{Chu2017,Satzinger2018,Kervinen2019}, and applications as quantum random access memories have  been proposed \cite{Hann2019}. 

Here, we develop a  multimode quantum acoustic device by integrating a superconducting quantum interference device (SQUID) into one of the Bragg reflectors of a SAW resonator. The SQUID inductance modulates the reflectivity of a unit cell in the mirror and hence the effective length of the resonator. Due to the narrow free spectral range, the SQUID reflector gives rise to coupling of more than 20 modes. We exploit this coupling to generate two-mode squeezed states between phonons in different SAW modes with a parametric drive. Extending the pump scheme to four tones, we demonstrate multipartite entanglement between four acoustic modes. Our results suggest this quantum acoustic platform can be used to create highly entangled multimode states for CV quantum computing.

\section{Device design and setup}

The SAW resonator, shown in Fig.~\ref{fig: schematic}, is defined by two reflectors with the leading edges separated by a distance of $\SI{600}{\upmu m}$. The reflector on the left hand side has 1200 fingers all shorted together. On the right hand side, the reflector has an interdigitated structure, where fingers are connected to either the top or bottom electrode in an alternating pattern. The top and bottom electrodes each have $N_p= 500$ fingers with an overlap of $W=\SI{100}{\upmu m}$, and are connected via a SQUID. An interdigitated transducer (IDT) centered between the reflectors provides a single port to the resonator. The port IDT has 75 periods and a double-finger structure to suppress mechanical reflections \cite{Datta1986}. An on-chip fluxline is used to apply an RF flux through the SQUID. The IDT and reflectors are fabricated from aluminium on a gallium arsenide substrate. Due to the piezoelectric coupling, the SAW field inside the resonator induces an electric potential difference between the top and bottom electrodes, generating currents through the SQUID.

Configurations where an interdigitated reflector is shunted by a variable load impedance have been used for SAW-based sensors \cite{Genji2014}. Here, the SQUID impedance provides a means of flux tuning the SAW resonator, as well as a cross-Kerr interaction between the modes. Integrating the SQUID makes the device a kind of acoustic analogue to the superconducting cavity-based Josephson parametric amplifier \cite{Roy2016}, where the short wavelength of sound allows for a much denser mode spacing than in the purely electromagnetic case. The SQUID reflector is equivalent to a dispersively coupled nonlinear resonator, as the interdigitated fingers give rise to a large capacitance connected in parallel with the SQUID inductance. We use this model to explain the effect of parametric modulation in this system.

\begin{figure}
	\centering
	\includegraphics[scale=0.75]{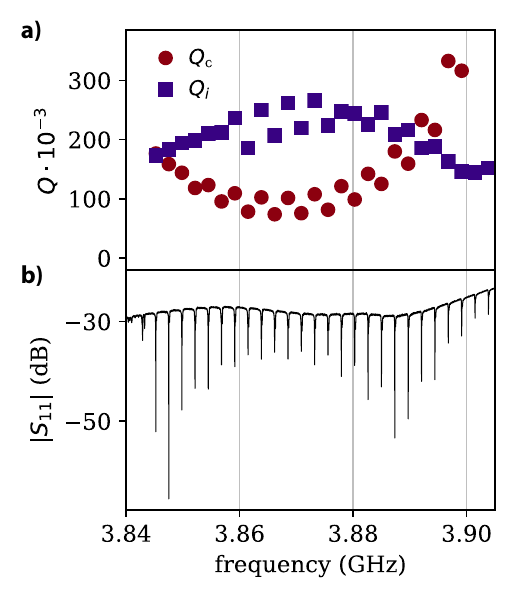}
	\caption{\textbf{Resonator characterization}. \textbf{a)} External and internal quality factors for each mode in the comb. Modes near the center of the 26-mode comb are overcoupled ($Q_c<Q_i$) and have internal quality factors $Q_i>10^5$ at high power. \textbf{b)} Reflection coefficient measured from the IDT with a vector network analyzer. The alternating even and odd mode pattern as well as the frequency dependence of the IDT response and mirror stopband are visible in the magnitude of resonance dips.}
	\label{fig: VNAscan}
\end{figure}

The mode structure of the resonator is shown in Fig.~\ref{fig: VNAscan}. While the IDT is centered with respect to the leading edge of each reflector, the broken symmetry due to the SQUID allows coupling to both odd and even modes with a free spectral range of $FSR = \SI{2.3}{MHz}$. The alternating pattern of even and odd modes is apparent in the external and internal quality factors $Q_c, Q_i$ extracted from fits to reflection measurements. The IDT couples more effectively to the even modes, resulting in a lower  $Q_c$. The frequency dependence of the IDT and mirrors provide a bandwidth of around \SI{40}{MHz} around the IDT center frequency where SAW modes are overcoupled. 

The narrow free spectral range of the resonator allows for simultaneous measurement of the response at multiple resonances all multiplexed in a single  channel. For this measurement we use a digital microwave measurement platform \cite{IMPwebsite} to directly digitally synthesize and measure signals at multiple frequencies simultaneously without analog mixers for frequency conversion.

\section{Coupled parametric resonator interaction}
The electromagnetic mode of the mirror has a frequency $\omega_{LC}$ which is parametrically modulated in time. The coupling to the SAW modes gives rise to the effective Hamiltonian (Appendix \ref{appendix: model})
\begin{equation}
\begin{split}
H_{\text{eff}} &= \sum_j \hbar \tilde{\omega}_j b_j^\dagger b_j + s(t)\sum_{j,k}\Tilde{g}_j\Tilde{g}_k\left(b_j - b_j^\dagger\right)\left(b_k - b_k^\dagger\right).
\end{split}
\label{eq: Hamiltonian}
\end{equation}
Here $b_j$ and $b_j^\dagger$ are the ladder operators for the SAW modes, while $s(t)$ is a time-dependent factor determined by the flux pump amplitude and frequency. Assuming a uniform vacuum coupling strength $g$, the effective coupling rates $\Tilde{g}_j$ depend on the mirror and SAW mode resonance frequencies as
\begin{align}
 \Tilde{g}_j &= \frac{2 g\omega_j}{\omega_j^2-\omega^2_{LC} } \label{eq: g_tilde}.
 \end{align}
The bare SAW frequencies $\omega_j$ are renormalized due to the interaction to
\begin{equation}
    \tilde{\omega}_j=\omega_j- g\tilde{g}_j\frac{\omega_{LC}}{\omega_j}.
\end{equation}

The second term in Eq. \eqref{eq: Hamiltonian} contains both beamsplitter and two-mode squeezing interactions. Depending on the modulation frequency of $s(t)$, either interaction can be selected. A beamsplitting interaction may be implemented by a parametric drive close to the difference frequency of the SAW modes. Modulating near the sum frequency will induce two-mode squeezing between pairs of modes. In our experiment we modulate the magnetic flux through the SQUID without DC flux bias at the frequency of a SAW mode. The parabolic dependence of the frequency on flux implies the electromagnetic mirror mode, and hence the effective coupling rate, are modulated at twice the pump frequency $s(t)\Tilde{g}_j\Tilde{g}_k \sim (\cos2\omega_pt + 1)\Tilde{g}_j\Tilde{g}_k$. This gives rise to two-mode squeezing as two photons from the pump tone are converted to one phonon each in modes symmetric around the pump.

\section{Two-mode squeezing}
\begin{figure}
	\centering
	\includegraphics[scale=0.66]{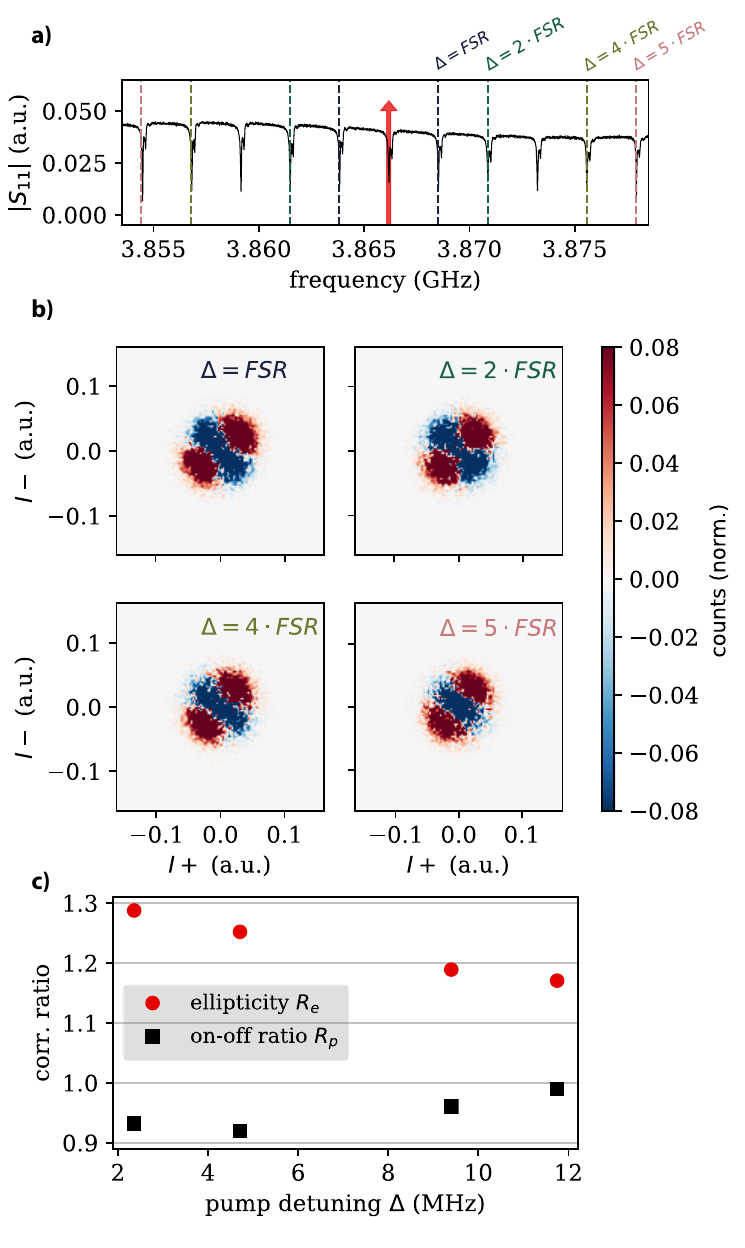}
	\caption{\textbf{Two-mode squeezing}. \textbf{a)} Measurement configuration where the red arrow indicates the pump frequency. Dashed lines indicate probe frequencies for collecting quadrature noise data. \textbf{b)} Histograms of two-mode correlation data where the measurement with the pump turned off has been subtracted. The label $I+$ ($I-$) denotes the $I$ quadrature of modes at frequencies above (below) the pump frequency. The label $\Delta$ indicates the detuning of the modes from the pump. The histograms for all two-mode quadrature combinations for the $\Delta=FSR$ case are shown in appendix~G. \textbf{c)} Squeezing ellipticity and ratio of standard deviations of the noise in the squeezed quadrature relative to the pump-off case. The quantities are plotted as a function of detuning of the measured mode pair from the pump. All modes show squeezing below the pump off level, and the error bars are smaller than the plot markers.}
	\label{fig: twomode_squeezing}
\end{figure}

To observe two-mode squeezing of the SAW field, we apply a parametric pump via the on-chip fluxline at the frequency of a SAW mode $f_p$. We measure the output field from the IDT at SAW mode frequencies symmetrically around the pump $f_{i,\pm}$, such that $2f_p -f_{i,-} - f_{i,+} =0$. The frequency configuration of the measurement is illustrated in Fig.~\ref{fig: twomode_squeezing}a. To characterize the correlations, we obtain reference histograms of the $I$ and $Q$ quadratures with the pump turned off. To minimize the effect of slow drift in the experimental setup, the pump output is switched on and off at a rate of \SI{2}{Hz}. The output signal is amplified using a travelling-wave parametric amplifier \cite{Macklin2015}. Data are collected over approximately 7 hours (3.5 hours each with the pump on and off). A quadrature rotation is applied to the measured data to compensate for slow phase drift in the experiment.

In Fig.~\ref{fig: twomode_squeezing}b we show subtracted quadrature histograms measured simultaneously in four pairs of SAW modes. Histograms are generated from $N=1.25 \cdot 10^6$ points measured in each mode with the pump on (off). The unsqueezed histograms obtained with the pump off are then subtracted from those produced with the pump on. We observe squeezing below the pump off level in all four mode pairs, extending the two-mode squeezing effect to SAW phonon fields. The ellipticity, defined as the ratio of the squeezed and anti-squeezed axes 
\begin{equation}
    R_e = \frac{\sigma_\mathrm{max}}{\sigma_\mathrm{min}} = \frac{\sqrt{\left\langle \left(I_+ +I_-\right)^2\right\rangle}}{\sqrt{\left\langle \left(I_+ - I_-\right)^2\right\rangle}}
\end{equation} 
is well above unity for all four mode pairs and shows a diminishing trend with increased detuning. The ratio of the standard deviation in the squeezed quadrature to the pump off case, given by
\begin{equation}
    R_p =  \frac{\sigma_\mathrm{min}}{\sigma_\mathrm{off}} = \frac{\sqrt{\left\langle \left(I_+ -I_-\right)^2\right\rangle}}{\sqrt{\left\langle \left(I_\mathrm{off}\right)^2\right\rangle}},
\end{equation}
has a value $R_p<1$ across the four mode pairs. The correlation ratios are plotted as a function of pump-probe detuning in Fig.~\ref{fig: twomode_squeezing}c. As expected, no correlations are observed outside the two-mode squeezed pairs. From our analysis of mode correlations (see appendix \ref{appendix: tms estimate}) we estimate that the squeezing is below the vacuum level, if the modes are cooled below an effective temperature of 80 mK. Even if our device is not perfectly thermalized to the \SI{10}{mK} cryostat temperature, it is unlikely the effective phonon temperature should exceed this bound, leading to strong indication that we have observed squeezing below the vacuum in the phonon field.

The ability to generate two-mode squeezing with a single pump tone is an important step towards multimode entanglement, as this can be achieved using multimode squeezing \cite{Menicucci2007}. In the next section we present such an experiment with a multitone pump and calibrated measurement chain.

\section{Multimode entanglement}
\label{sec: multimod}
Following the two-mode squeezing measurement, we develop the experiment further to observe multipartite entanglement involving more SAW modes. For this purpose we use another device with reduced cross-talk between the pump line and IDT described in more detail in Appendix~\ref{appendix: device B}. With a multi-tone modulation, the Hamiltonian of Eq.~\ref{eq: Hamiltonian} can provide coupling between any pair of modes in the resonator. Instead of single pump tone, we now apply a regularly-spaced comb of pump frequencies containing up to four tones. Due to the slight deviation from equidistance in the SAW modes the resulting entanglement is restricted to a set of modes in the vicinity of the pumps. The sharp mode structure also implies the amplitude and phase of correlations are sensitive to the pump comb settings. This is apparent in the scattering response shown in Appendix \ref{appendix: scattering}.

For the two-mode squeezing measurement data we subtract the noise measured with the pump off. In order to establish multipartite entanglement we instead perform a calibration of the gain (Appendix~\ref{appendix: calibration}) and added noise of the measurement amplification chain. The calibration is based on Planck spectroscopy \cite{Mariantoni2010} and provides an estimate of the power level corresponding to vacuum fluctuations in the SAW modes, allowing us to test the measurement data for continuous-variable entanglement.

The Gaussian state of the probe modes is characterized by the quadrature covariance matrix. Drift and noise in the measurement setup can diminish mode correlations and render averaged covariance matrices unphysical. To mitigate this problem, we divide the 2.5-minute measurement into two-second intervals and perform a reconstruction \cite{Shchukin_2016} to ensure a physical state and test for entanglement on each interval separately.

\begin{table}
\caption{Significance of detected entanglement for all bipartitions using the estimate of Eqs.~\ref{eq:SvL criteria bipart}-\ref{eq: entanglement}. The highest significance is obtained for partitions separating modes 3 and 4.} 
 \begin{tabular}{ c c } 

 bipartition & $\Sigma_w$ \\
 \hline\hline
 $\{1\} : \{2, 3, 4\}$ & $-13.7$ \\ 
 $\{2\} : \{1, 3, 4\}$ & $-2.4$ \\
 $\{1,2\} : \{3, 4\}$ & $-2.9$ \\
 $\{3\} : \{1, 2, 4\}$ & $-70.6$ \\
 $\{1,3\} : \{2, 4\}$ & $-74.1$ \\
 $\{2,3\} : \{1, 4\}$ & $-69.3$ \\
 $\{1, 2, 3\} : \{4\}$ & $-75.7$ \\

\end{tabular}
\label{tab: significance}
\end{table}
\begin{figure}
	\centering
	\includegraphics[scale=0.66]{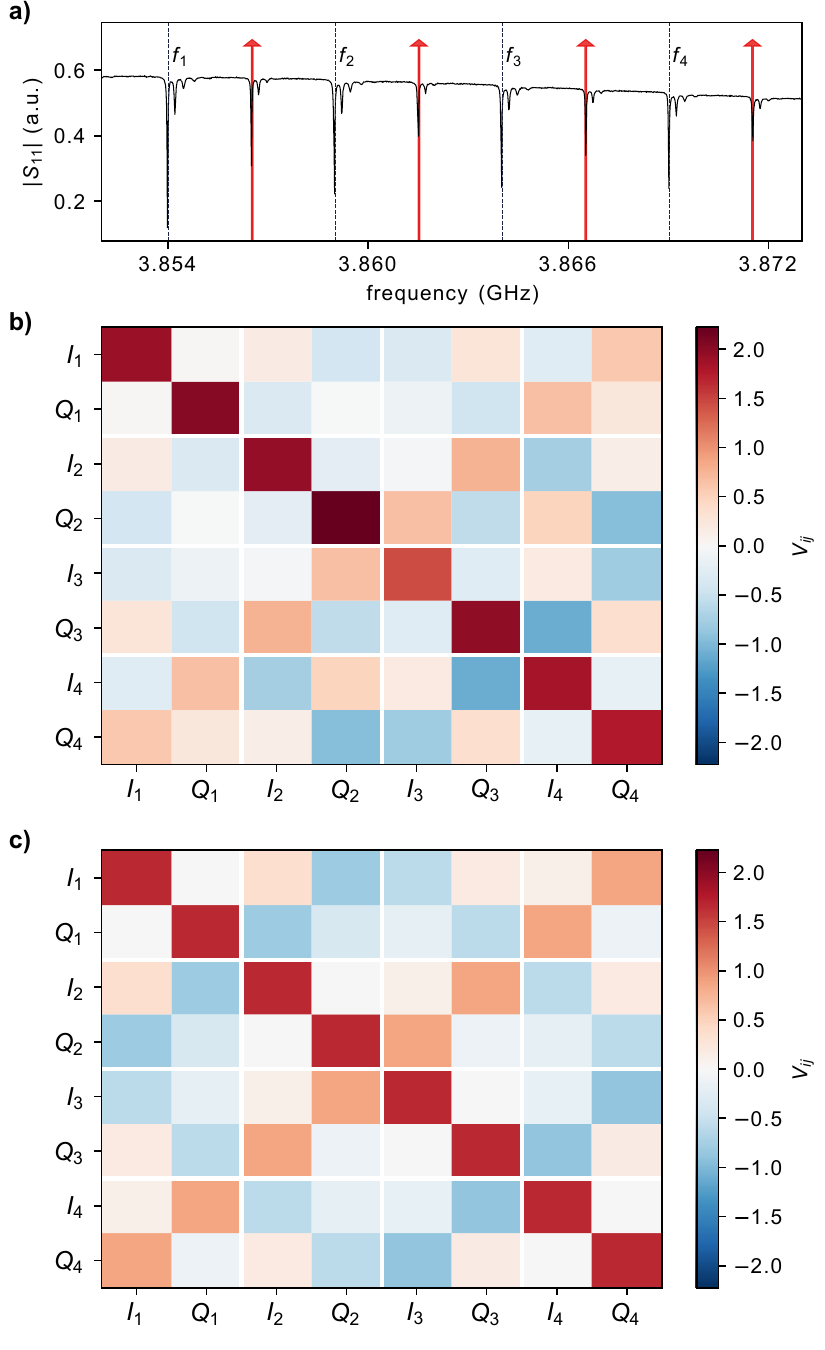}
	\caption{\textbf{Multimode covariance matrix}. \textbf{a)} Measurement configuration where the solid red arrows indicate the pump frequencies. Dashed lines indicate probe frequencies for collecting quadrature noise data. \textbf{b)} Four-mode quadrature correlation matrix generated from  data measured with the pump scheme shown in \textbf{a}. \textbf{c)} Theoretical covariance matrix. The analyzed measurement configuration corresponds approximately to that of the experimental data. While in the calculation the pump strength and loss rates are uniform across the modes, the relative pump phases have been adjusted to reproduce qualitatively the features of the measured result shown in \textbf{b}.}
	\label{fig: twopump covar}
\end{figure}
The pump and measurement configuration for four modes using four pump tones is shown in Fig.~\ref{fig: twopump covar}a. Figure~\ref{fig: twopump covar}b shows the quadrature covariance matrix obtained in this measurement. Using the calibration reference, the measured amplitudes are scaled with the single photon energy and measurement bandwidth $\Delta_{BW}$ as
\begin{equation}
V_{ij}=\frac{\left\langle A_i A_j \right\rangle}{\frac{1}{2} Z_0\hbar \sqrt{\omega_i \omega_j} \Delta_{BW} }
\end{equation}
where $A \in \{I,Q\}$. With this scaling the vacuum state is given by the identity matrix. The measured mode set extends beyond the four modes analyzed here, but we are not able to recover physical covariance matrices for all modes from our calibration of the amplifier gain and added noise.

The histograms of the output quadratures for the $f_k$ modes are all measured in parallel. In this case the correlations are not restricted to pairwise two-mode squeezing, but all modes are mutually correlated. As shown in Fig.~\ref{fig: twopump covar}c, the correlation features are qualitatively captured by our theoretical model presented in Appendix \ref{appendix: theory cov matrix}.

The multimode correlated state can be analyzed for entanglement. We evaluate the entanglement using a variant of negativity of partial transpositions \cite{Simon2000} developed in \cite{Shchukin2015}. This test relies on violating the inequality 
\begin{align}
   \mathcal{E} =& \text{Tr}\left[V^{II} (\mathbf{h} \otimes \mathbf{h}) \right] + \text{Tr}\left[V^{QQ} (\mathbf{g} \otimes \mathbf{g}) \right] \nonumber
   \\ 
   &- 2 |\left\langle h_\mathcal{I}, g_\mathcal{I} \right\rangle| - 2 |\left\langle h_\mathcal{J}, g_\mathcal{J} \right\rangle| \geq 0
   \label{eq:SvL criteria bipart}
\end{align}
which holds for separable states.

In computing the quantity $\mathcal{E}$, the covariance matrix is rotated to eliminate correlations between $I$ and $Q$ quadratures and $V^{II}$ ($V^{QQ}$) denotes the submatrix containing the $I-I$ ($Q-Q$) correlations. The vectors $\mathbf{h}$ and $\mathbf{g}$ are real-valued with lengths equal to the number of modes. The subscripted terms indicate elements (and their corresponding modes) of $\mathbf{h}$ and $\mathbf{g}$ belonging to the bipartition subsets $\mathcal{I}$ and $\mathcal{J}$. One should consider $\mathbf{h}$ and $\mathbf{g}$ as the coefficients of a general test operator acting as our entanglement witness \cite{Gerke2015}. We are free to optimize $\mathbf{h}$ and $\mathbf{g}$ to maximize any violation of the inequality Eq.~\eqref{eq:SvL criteria bipart} for a given bipartition $\mathcal{I}$ and $\mathcal{J}$.
The entanglement measure is then obtained as a weighted mean over all 75 intervals within the full integration time. We estimate the standard deviation in the measurement and express entanglement in terms of the significance $\Sigma_w$, given by
\begin{equation}
    \Sigma_w = \frac{ \mathcal{E} }{\sigma}.
    \label{eq: entanglement}
\end{equation}
The uncertainty $\sigma$ is calculated as
\begin{equation}
    \sigma = \sqrt{ \sum_{ij} \left( {\sigma}^2_{ij} h^2_i h^2_j + { \sigma}^2_{ij} g^2_i g^2_j \right) }. \label{eq:error entanglement}
\end{equation}
where the matrix elements $\sigma_{ij}$ are obtained by error propagation accounting for the uncertainty in the calibration gain and noise parameters as well as noise in the measurement (appendix \ref{appendix: calibration}).

The entanglement significance is computed for all bipartitions of the four-mode set. As shown in Table~\ref{tab: significance}, all bipartitions yield entanglement by at least 2.4 standard deviations. Negativity of the entanglement test for \emph{all} bipartitions is a signature of full multipartite entanglement \cite{Weedbrook2012}. We observe that the entanglement significance is substantially higher for bipartitions where the modes 3 and 4 appear in separate sets. This is due to the imperfect alignment of the equidistant pump comb with the mode structure. In the measured covariance matrix in Fig.~\ref{fig: twopump covar}b, this leads to stronger correlations involving modes 3 and 4. More uniform correlations and enhanced entanglement significance can be obtained by optimizing the pump settings.

Tailoring the digitally synthesized microwave frequency pump spectrum is also a way to obtain different entanglement structures in this setup. The square lattice is one example of a cluster state that can be used to to implement universal quantum computation and can be generated from the Hamiltonian of Eq.~\ref{eq: Hamiltonian} \cite{Menicucci2007}. The digital control of the amplitude and phase of each pump tone also enables extending the measurement scheme to observe multipartite entangled states with sizes approaching the number of modes in the SAW resonator. Beyond the straight-forward generation of particular target cluster states, technical (and theoretical) challenges remain to overcome errors due to the finite squeezing and achieve fault-tolerance \cite{Menicucci2014}.

\section{Conclusions}

We have demonstrated two-mode squeezing in a surface acoustic wave resonator, likely below the phononic vacuum level. Extending this scheme to a multitone pump spectrum and calibrated measurement chain, we observed fully inseparable multipartite entanglement of four resonator modes. The dense mode structure of the resonator enables multiplexing all modes to one measurement channel without analog frequency conversion. The small on-chip footprint of our device ($<\SI{0.2}{mm^2}$) further contributes to scalability. 

For the measurements presented here, the pump strengths are similar to the loss rates. This limits the amount of squeezing and correlations that can be induced. To enable stronger pumping and enhance the entanglement generation, a 3-wave mixing scheme could be adopted where the pumping occurs at around twice the mode frequencies. For the same drive amplitude, this yields stronger pumping as well as less effect of saturation in the parametric amplifier. A prospect for further development is using superconducting qubits to implement non-Gaussian operations on the resonator state such as the addition or subtraction of single phonons. Non-Gaussianity is important to many applications \cite{Ra2019} and qubit-controlled operations on the resonator state are more readily implemented in a circuit QED setting than optical experiments where nonlinearities are typically weaker. As a step in this direction, the device used to measure multimode correlations has an integrated transmon qubit, although it was not used for this experiment.

Another promising application for this device is in quantum simulation using the resonator modes as lattice sites in a synthetic dimension. Modulating the reflector SQUID at a frequency corresponding to the the free spectral range will induce nearest-neighbour hopping of phonons, giving rise to an effective lattice Hamiltonian in a hardware-efficient way. 

\section{Acknowledgements}
We acknowledge IARPA and Lincoln Labs for providing the TWPA used in this experiment. We are grateful to G. Ferrini, I Strandberg and F. Quijandria for fruitful discussions. R.B., M.O.T., and D.B.H. are part owners of the company Intermodulation Products AB, which produces the digital multifrequency lock-in amplifier used in this experiment. This work was supported by the Knut and Alice Wallenberg foundation through the Wallenberg Center for Quantum Technology (WACQT), and the Swedish Research Council, VR.

%


\appendix

\section{Coupling strength estimate}

The vacuum coupling strength between the LC mode of the SQUID mirror and the SAW modes is given by the overlap of the zero-point voltage fluctuations of the SAW mode with the charge fluctuations on the mirror fingers \cite{Moores2018}. As any acoustic mode that is confined in the resonator is necessarily efficiently reflected by the mirror, we make the simplifying assumption that the SAW wavelength matches the mirror period for all modes. The amplitude of the voltage zero-point fluctuations is given by
\begin{equation}
    \phi_0 = \frac{e_{14}}{\epsilon}\sqrt{\frac{\hbar}{2\rho v_\mathrm{SAW}A}}.
\end{equation}
The piezoelectric coefficient $e_{14}$ and the dielectric constant $\epsilon$, as well as the substrate density $\rho$ and SAW velocity $v_\mathrm{SAW}$ are material parameters, while $A$ denotes the effective mode area. The charge fluctuations across the mirror fingers are
\begin{equation}
    Q_0 = 2e\beta \left(\frac{E_L}{32E_C}\right)^{1/4}
\end{equation}
where $E_L = (\Phi_0/2\pi)^2/L_J$ is the characteristic inductive energy, and $E_C=e^2/(2C)$ sets the charging energy scale. The capacitance ratio $\beta$ indicates the ratio of the mirror capacitance seen by the SAW modes to the total mirror capacitance. Because the SAW field decays exponentially into the mirror with a penetration depth $L_p$, this ratio is approximately given by $\beta=L_p/L_m$, where $L_m$ is the total length of the mirror. This yields an approximate coupling strength
\begin{equation}
\hbar g = \phi_0 Q_0 = e\frac{e_{14}}{\epsilon}\frac{L_p}{L_m} \left(\frac{E_L}{8E_C}\right)^{1/4}\sqrt{\frac{\hbar}{\rho v_\mathrm{SAW}A}}.
\label{eq: coupling strength}
\end{equation}
With literature values for the material parameters \cite{Simon1996, Aref2016} and the penetration depth $L_P$ estimated from the measured free spectral range, we obtain $g\approx 2\pi \cdot \SI{1.6}{MHz}$. This weak vacuum coupling strength implies the dispersive interaction between SAW modes and the mirror are small compared to the linewidth, and parametric excitation thus relies on strong pumping.

\section{Device B}
\label{appendix: device B}
The device used for the multimode entanglement experiments of Sec.~\ref{sec: multimod} has slight design variations. The mirror edge separation is $L_e=\SI{560}{\upmu m}$. {To enhance the coupling of the electromagnetic mirror mode to SAW, the number of finger pairs in the mirror has been reduced to $N_p=275$.} The estimated vacuum coupling (Eq.~\ref{eq: coupling strength}) is $g=2\pi \cdot \SI{1.6}{MHz}$. 

\begin{figure}[htbp]
	\centering
	\includegraphics[scale=0.22]{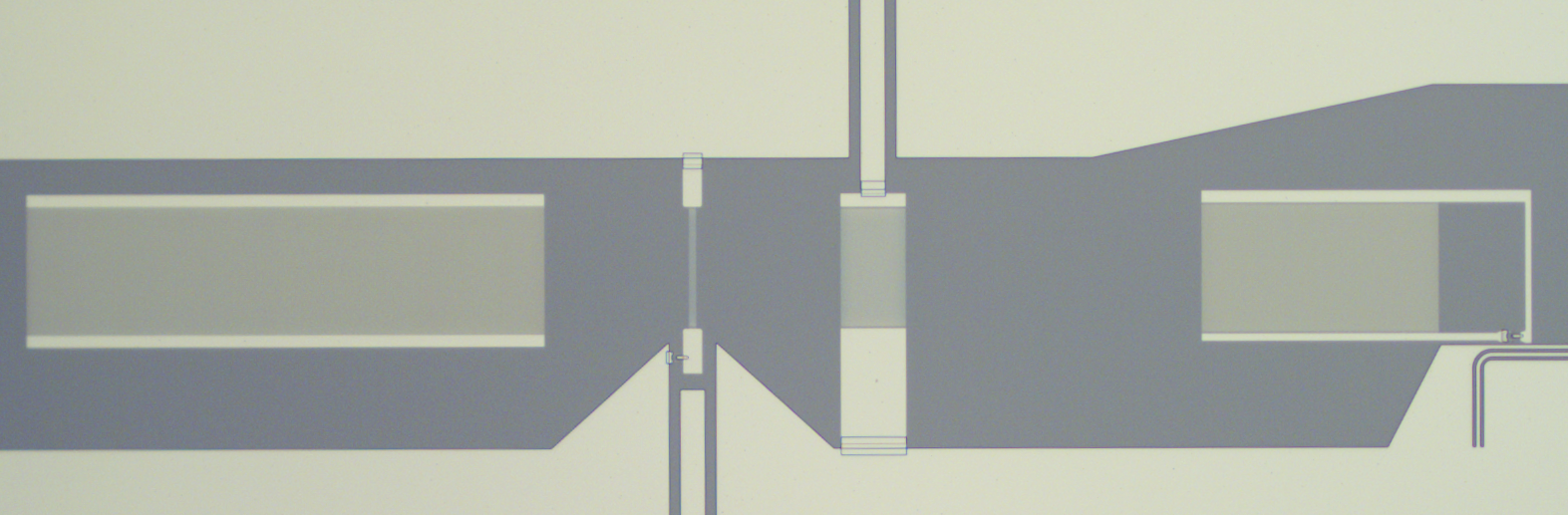}
	\caption{\textbf{Device B.} The sample used for the multimode entanglement measurements presented in Sec.~\ref{sec: multimod}. }
	\label{fig: device B image}
\end{figure}

The total capacitance is $C=\SI{3.3}{pF}$ and the SQUID shunting the two electrodes has a total critical current of $I=\SI{190}{nA}$ ($L_J = \SI{1.7}{nH}$). The associated $LC$ mode has a frequency of $\omega_{LC}=2\pi \cdot \SI{2.1}{GHz}$. To enable further operations on the resonator state, a transmon qubit has been integrated between the port IDT and left hand mirror. While not used in the present experiment, qubit operations could be relevant to creating non-gaussian SAW states. A microscope image of the device is shown in Fig.~\ref{fig: device B image}.

\section{Experimental details}
\label{appendix: setup}

\begin{figure}[htbp]
	\centering
	\includegraphics[scale=0.35]{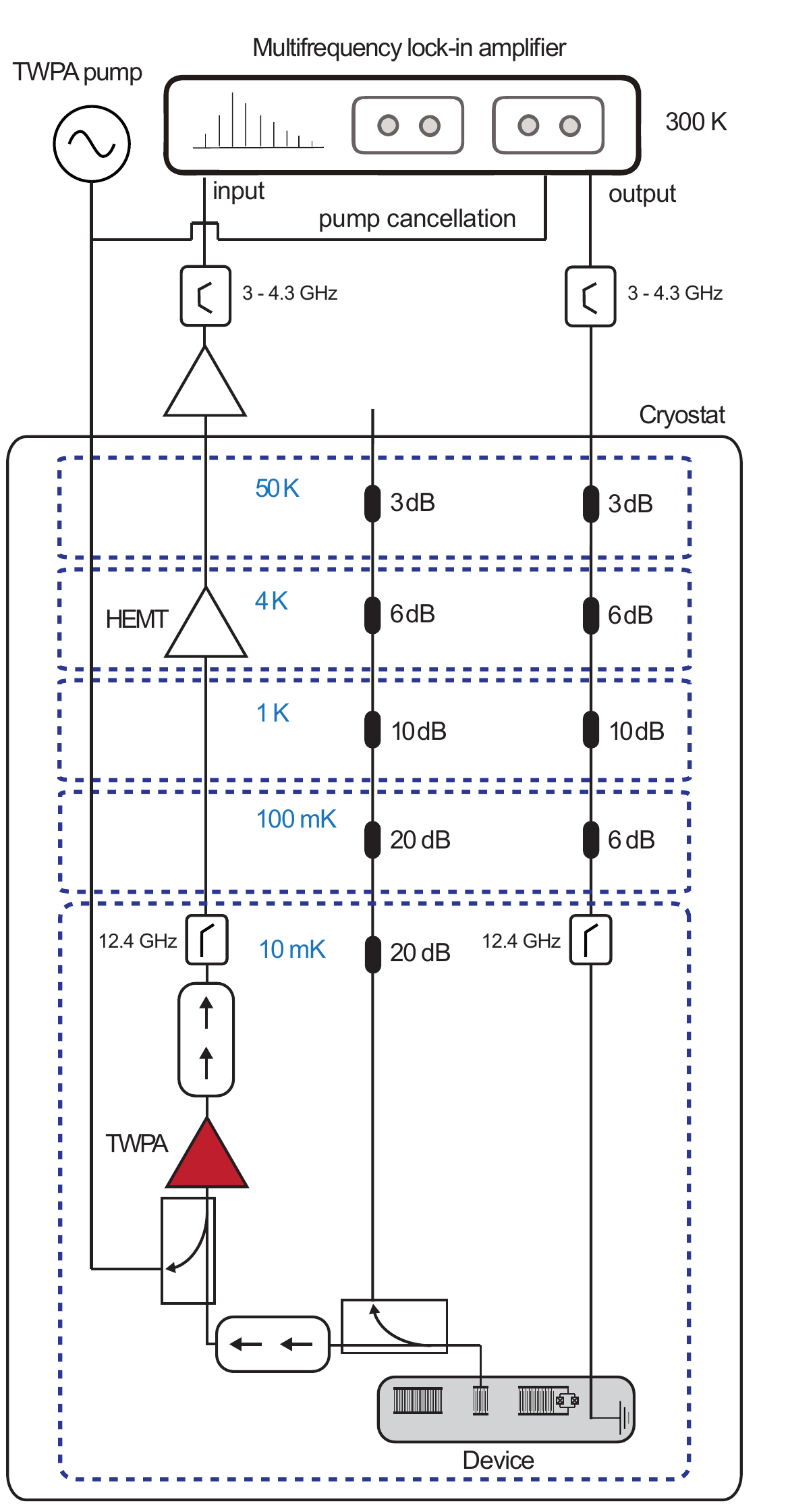}
	\caption{\textbf{Schematic of the measurement setup.} {The signal output and input have band-pass filters to suppress harmonics and spurious tones. To avoid crosstalk from the flux pump line saturating the TWPA, we use a destructively interfering cancellation tone at the pump frequency. This cancellation tone, only used in the two-mode squeezing experiment, is output from a separate channel and combined with the TWPA pump. The input line to the IDT is used for characterizing the resonator reflection.}}
	\label{fig: setup}
\end{figure}

A schematic illustrating the measurement setup is shown in Fig. \ref{fig: setup}.
The digital microwave measurement platform has 8 channels of high speed digital-to-analog (DAC) and analog-to-digital converters (ADC) serviced by a large field programmable gate array (FPGA), all synchronized to one stable clock \cite{IMPwebsite}. With this setup we are able to digitally synthesize drive signals and digitize response signals in the band 2-4 GHz, without analog IQ mixers. We set the sampling clock at 4 GSamples/s for the ADCs, and at 5 GSamples/s for the DACs, resulting in a Nyquist frequency of 2 GHz and 2.5 GHz respectively. The SAW cavity modes are designed to fall in the band 3.8 - 3.9 GHz, within the second Nyquist zone of the converters. Tones outside the second Nyquist zone are removed with external bandpass filters.

\section{Parametric coupling model}
\label{appendix: model}
In this model we consider the coupling induced between linear SAW modes by the common interaction with the \emph{LC} resonance of the mirror under parametric modulation. If we indicate the \textit{LC} mode and SAW modes by the $a$ and $b_j$ operators, respectively, the total Hamiltonian is a sum of three terms, $H=H_{0} + V + D(t)$ given by
\begin{align}
    H_0/\hbar &= \omega_{LC} a^\dagger a + \sum_j \omega_j b_j^\dagger b_j, \\
    V/\hbar &= i\sum_j g \left(a - a^\dagger\right)\left(b_j + b_j^\dagger \right), \\
    D(t)/\hbar &= -\frac{1}{2} \omega_{LC} \left( 1 - \cos{ \frac{\pi \Phi(t)}{\Phi_0} } \right) (a + a^\dagger)^2,
\end{align}
where the driving term $D(t)$ is the time-dependent part of the Josephson energy due to the flux pump $\Phi(t)$. 

If $|\Phi(t)| \ll \Phi_0$, we can expand the driving term $D(t)$ to second order in $\Phi(t)$
\begin{align}
    D(t) &\approx -\frac{1}{4}\hbar \omega_{LC} \left( \frac{\pi \Phi(t)}{\Phi_0} \right)^2 (a + a^\dagger)^2. \label{eq: general drive term to second order}
\end{align}
In the case for a single flux pump, the time-dependence is described by a sinusoidal function $\Phi(t) = \Phi_{AC}\cos(\omega_p t + \theta)$. 
Inserting this into Eq. \eqref{eq: general drive term to second order}, we arrive at the expression of the drive term for a single flux pump at $\omega_p$
\begin{align}
    D(t) &= - 2 d \cos^2{\left( \omega_p t +\theta \right)}(a+a^\dagger)^2 \nonumber \\
    &= -d \left( \cos{ \left(2\omega_p t + 2\theta \right)} + 1 \right)(a+a^\dagger)^2
    \label{eq: drive term single pump}
\end{align}
where $d = \hbar\omega_{LC}(\pi \Phi_{AC}/2\Phi_0)^2/2$ is the effective pump amplitude. For multiple flux pumps at different frequencies, $\Phi(t)$ is instead a superposition of sinusoidal functions.

{In the limit of weak interaction between the electromagnetic mirror mode and SAW}, $g \ll |\omega_i - \omega_{LC}|$, we perturbatively expand the Hamiltonian by applying the Schrieffer-Wolff transformation \cite{Schrieffer1966}
\begin{align}
    H' = e^{-S} H e^{S} \approx H_0 + \frac{1}{2}[S, V] + e^{-S}D(t)e^{S} \label{eq: SW transformation}
\end{align}
where
\begin{align}
    S = \sum_j &\left[ \frac{i g}{ \omega_{LC} - \omega_j } \left( a^\dagger b_j + a b_j^\dagger \right) + \frac{i g}{ \omega_{LC}  + \omega_j} \left(a b_j + a^\dagger b_j^\dagger \right) \right]
\end{align}
and we use the form of the drive term $D(t)$ given by Eq. \ref{eq: drive term single pump}.
The commutator $[S, V]$ is therefore
\begin{align}
    [S, V] &= -g \hbar \left( \sum_j \tilde{g}_j (a-a^\dagger)^2 - \sum_{j,k}\bar{g}_j(b_j + b_j^\dagger)(b_k + b_k^\dagger) \right).
\end{align}
where we introduce the effective coupling rates $\tilde{g}_j$ and $\bar{g}_j$:
\begin{align}
    \tilde{g}_j &= \frac{2 g \omega_j}{\omega_j^2 - \omega_{LC}^2}, \\
    \bar{g}_j &= \frac{-2 g \omega_{LC}}{\omega_j^2 - \omega_{LC}^2}.
\end{align}

\begin{widetext}
We also treat the pump-dependent term $e^{-S}D(t)e^{S}$ perturbatively, by expanding up to second order in $S$ according to
\begin{align}
    e^{-S}D(t)e^{S} &\approx D(t) - [S, D(t)] + \frac{1}{2}[S, [S, D(t)]].
\end{align}
The commutators are calculated to be
\begin{align}
    [S, D(t)] &= -2 d \left( \cos{(2\omega_p t + 2\theta)} + 1 \right)\sum_j i \tilde{g}_j (a+a^\dag)(b_j-b_j^\dag), \label{eq: mirror-saw_coupling} \\
    [S, [S, D(t)]] &= - 2 d \left( \cos{(2\omega_p t + 2\theta)} + 1 \right) \left( -\sum_{j,k} \tilde{g}_j \tilde{g}_k ( b_j-b_j^\dag)(b_k-b_k^\dag) + \sum_j\bar{g}_j\tilde{g}_j( a+a^\dag)^2 \right) \label{eq: SAW_squeezing_term},
\end{align}
where we have neglected an unimportant constant term.

To summarize, we write down the Hamiltonian $H'$
\begin{gather}
    H' = H_0 + D(t) - \frac{g \hbar}{2} \left( \sum_j \tilde{g}_j (a-a^\dagger)^2 - \sum_{j,k}\bar{g}_j(b_j + b_j^\dagger)(b_k + b_k^\dagger) \right) + 2 d \left( \cos{(2\omega_p t + 2\theta)} + 1 \right) \sum_j i \tilde{g}_j (a+a^\dag)(b_j-b_j^\dag) \nonumber \\
    + d \left( \cos{(2\omega_p t + 2\theta)} + 1 \right) \left( \sum_{j, k} \tilde{g}_j \tilde{g}_k ( b_j-b_j^\dag)(b_k-b_k^\dag) - \sum_j\bar{g}_j\tilde{g}_j ( a+a^\dag)^2 \right),
\end{gather}
where the last term produces squeezing and beamsplitter interactions.
Finally, we perform a resonance approximation by dropping all rapidly oscillating terms. This leaves us with the Hamiltonian $\tilde{H}$
\begin{gather}
    \tilde{H} = \hbar \tilde{\omega}_{LC} a^\dagger a + \hbar\sum_j \left( \tilde{\omega}_j - 2d \tilde{g}^2 \right)b_j^\dagger b_j
    + \sum_{j}\sum_{k = 2 p - j}\frac{d \tilde{g}_j \tilde{g}_k}{2} \left( e^{2i(\omega_p t + \theta)} b_j b_k + e^{-2 i (\omega_p t + \theta)} b_j^\dagger b_k^\dagger \right) \label{eq: final_hamiltonian}
\end{gather}
where the final sum is only over SAW-modes $k$ satisfying the 4-wave mixing criterion $\omega_k = 2\omega_p - \omega_j$.
For multiple pumps, this would result in several 4-wave mixing criteria (one for each pump) and thus couple each SAW mode to more modes.
We also define the renormalized frequencies $\tilde{\omega}_{LC}$ and $\tilde{\omega}_j$ as
\begin{align}
    \tilde{\omega}_{LC} &= \omega_{LC} + g\sum_j\tilde{g}_j- \frac{2d}{\hbar} \left( 1 + \sum_{j}\bar{g}_j\tilde{g}_j \right), \\
    \tilde{\omega}_j &= \omega_j + g\bar{g}_j.
\end{align}
\end{widetext}

\section{Calculating the theoretical covariance matrix} \label{appendix: theory cov matrix}

Here we outline how to arrive at a covariance matrix from a system of Langevin equations. Calculating the Heisenberg equations of motion for the SAW modes $b$ using Hamiltonian $\tilde{H}$ Eq. \eqref{eq: final_hamiltonian}, we arrive at a system of equations describing multiple parametrically coupled modes. Assuming identical external couplings $\gamma$ and no internal losses, we arrive at \cite{Wustmann2017}
\begin{align}
   \dot{b}_j + i\left(\Tilde{\omega}_j - 2 d \tilde{g}_j^2 / \hbar  \right)b_j + \frac{\gamma}{2} b_j + i \sum_k\epsilon_{jk} b^\dagger_k = \sqrt{\gamma}b^{\text{in}}, \label{eq: time domain coupling eq}
\end{align}
where we define the complex parametric coupling $\epsilon_{jk}$ to be 
\begin{align}
    \epsilon_{jk} = \frac{d \tilde{g}_j \tilde{g}_k}{2 \hbar} e^{-2i(\omega_p t + \theta)}. 
\end{align}
The sum is made over all modes $k$ satisfying all 4-wave mixing criteria.

Since the FSR is small compared to $\tilde{\omega}_j$ and the {detuning between the SAW modes and the mirror electromagnetic mode,} we make the approximation $\tilde{g}_j \approx \tilde{g}_k$ which corresponds to the idealized case with identical parametric couplings. Eq. \eqref{eq: time domain coupling eq} is then instead
\begin{align}
    \dot{b}_j + i\left(\Tilde{\omega}_j - 4 |\epsilon|  \right)b_j + \frac{\gamma}{2} b_j + i \epsilon \sum_k b^\dagger_k = \sqrt{\gamma}b^{\text{in}}. \label{eq: heisenberg_eom_idealized}
\end{align}

Working in the frequency domain is more convenient. The Fourier transform of Eq. \eqref{eq: heisenberg_eom_idealized} is
\begin{align}
   -i\Delta_j b_j[\Omega_j] + i \epsilon \sum_k b^\dagger_j[\Omega_j] = \sqrt{\gamma}b^\text{in}[\Omega_j], \label{eq: freq domain coupling eq}
\end{align}
where $\Delta_j = \Omega_j - \Tilde{\omega}_n + 4|\epsilon| + i\gamma/2$. If the measurement frequency $\Omega_j = \Tilde{\omega}_j -  4|\epsilon|$, the expression simplifies to $\Delta = i \gamma/2$.

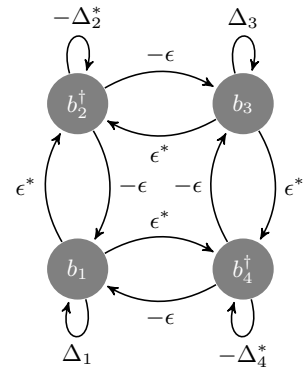
\begin{figure}[h]
    \centering

\begin{tikzpicture}[->,>=stealth',shorten >=1pt,auto,node distance=2.2cm,
                    semithick]
  \tikzstyle{every state}=[fill=gray,draw=none,text=white]

  \node[state]         (A)              {$b_1$};
  \node[state]         (B) [above of=A]  {$b_2^\dagger$};
  \node[state]         (C) [right of=B]  {$b_3$};
  \node[state]         (D) [below of=C]  {$b_4^\dagger$};

  \path (B) edge [bend left] node {$-\epsilon$} (A);
  \path (A) edge [bend left] node {$\epsilon^*$} (B);
  \path (B) edge [bend left] node {$-\epsilon$} (C);
  \path (C) edge [bend left] node {$\epsilon^*$} (B);
  \path (C) edge [bend left] node {$\epsilon^*$} (D);
  \path (D) edge [bend left] node {$-\epsilon$} (C);
  \path (D) edge [bend left] node {$-\epsilon$} (A);
  \path (A) edge [bend left] node {$\epsilon^*$} (D);
  
  \path (D) edge [loop below] node {$-\Delta_4^*$} (D);
  \path (C) edge [loop above] node {$\Delta_3$} (C);
  \path (B) edge [loop above] node {$-\Delta_2^*$} (B);
  \path (A) edge [loop below] node {$\Delta_1$} (A);
  
\end{tikzpicture}

    \caption{\textbf{Graph for multimode squeezing}. This graph represents the mode-couplings in Fig. \ref{fig: twopump covar} in the main text. For simplicity, we assume identical couplings.}
    \label{fig:example_graph}
\end{figure}

For many pumps and modes, the system of equations can be conveniently summarized by a complex weighted directed graph \cite{Ranzani2015}.
We consider the case for the multipartite entanglement result of Fig. \ref{fig: twopump covar} and draw the corresponding graph in Fig. \ref{fig:example_graph}. 

The graph illustrates the mode-couplings and can also be associated with a mode-coupling matrix $M$.
In the basis of $\Bar{b} = (b_1, ..., b_4, b^\dagger_1, ..., b^\dagger_4)$, the matrix $M$ is
\begin{align}
    M &= \begin{pmatrix}
    \Delta_1 & 0 & 0 & 0 & 0 & -\epsilon & 0 & -\epsilon\\
    0 & \Delta_2 & 0 & 0 & -\epsilon & 0 & -\epsilon & 0\\
    0 & 0 & \Delta_3 & 0 & 0 & -\epsilon & 0 & -\epsilon\\
    0 & 0 & 0 & \Delta_4 & -\epsilon & 0 & -\epsilon & 0\\
    0 & \epsilon^* & 0 & \epsilon^* & -\Delta_1^* & 0 & 0 & 0\\
    \epsilon^* & 0 & \epsilon^* & 0 & 0 & -\Delta_2^* & 0 & 0\\
    0 & \epsilon^* & 0 & \epsilon^* & 0 & 0 & -\Delta_3^* & 0\\
    \epsilon^* & 0 & \epsilon^* & 0 & 0 & 0 & 0 & -\Delta_4^*\\
    \end{pmatrix},
    \label{eq: M}
\end{align}
which provides a complete description of the frequency domain expression by $-i M\Bar{b} = \sqrt{\gamma}\Bar{b}^\text{in}$.

The covariance matrix $V$ can be obtained from $M$ via the scattering matrix $S$, given by
\begin{align}
    S &= i K M^{-1} K - I, \label{eq: scattering matrix}\\
    K &= \sqrt{\gamma}I.
\end{align}
The scattering matrix relates the incoming modes to the outgoing modes $\Bar{b}^{\text{out}} = S \Bar{b}^{\text{in}}$ according to input-output theory \cite{Gardiner1985, Wustmann2017, Ranzani2015}. The covariance matrix of the incoming noise modes $V^\text{in}$ then transforms as \cite{Weedbrook2012}
\begin{align}
    V^\text{out} = S_{IQ} V^\text{in} S^T_{IQ}
     \label{eq: V(S)}
\end{align}
where the subscript $IQ$ indicates the scattering matrix has been (linearly) transformed into the quadrature basis defined by $I= b + b^\dagger$, $Q = -i\left(b-b^\dagger\right)$. Any Gaussian state is fully characterized by $V^\text{out}$.

In the case of internal losses $\gamma^\text{int}$, we need to make some minor adjustments to the preceding method. The definition of $\Delta_j$ is adjusted to be $\Delta_j = \Omega_j - \Tilde{\omega}_j + 4|\epsilon| + i \gamma^{\text{tot}}/2$ with $\gamma^\text{tot} = \gamma + \gamma^\text{int}$. In addition, we introduce the diagonal matrix $K^{\text{int}} = \sqrt{\gamma^\text{int}} I$ which is used to define a scattering matrix for the loss channel as
\begin{align}
    S^\text{loss} &= i K M^{-1} K^{\text{int}}. \label{eq: scattering matrix w loss}
\end{align}
In the presence of internal losses, Eq. \eqref{eq: V(S)} is instead replaced by
\begin{align}
    V^\text{out} = S_{IQ} V^\text{in} S^T_{IQ} + S^\text{loss}_{IQ} V^\text{loss} \left(S^\text{loss}_{IQ}\right)^T.
    \label{eq: V(S) w losses}
\end{align}
The noise coming from the internal loss port is characterized by $V^\text{loss}$. Typically, it is assumed to be identical to the incoming noise port $V^\text{in} = V^\text{loss}$.

So far in this discussion, the pump modes have been ignored. These modes form a set of correlated modes which is separate from the probe modes and can therefore be ignored in the analysis. We also note that due to the restricted probe mode set, the highest-frequency pump tone used in the measurement does not contribute to the measured correlations.

A covariance matrix calculated theoretically from Eqs.~(\ref{eq: M}-\ref{eq: V(S) w losses}) is shown in Fig.~\ref{fig: twopump covar}c. We use a uniform pump strength of $|\epsilon | = \SI{30}{kHz}$ and equal external and internal loss rates of $\gamma=\gamma_\mathrm{ext}=\SI{20}{kHz}$. Although simplified, this configuration corresponds approximately to that of the multimode entanglement experiment and the theoretical covariance matrix qualitatively reproduces the features of the measured data shown in Fig. 4b.

\section{Amplifier Gain and added noise}

We model the effect of amplification on the covariance matrix according to \cite{Weedbrook2012}
\begin{equation}
    \tilde{V} = T V T + N 
    \label{eq: amp_transform}
\end{equation}
where $T = \sqrt{G} \, I$ and $N = (G - 1)(2n + 1) \, I$.
Our amplifier chain is characterized by an effective amplitude gain $\sqrt{G}$ and effective added mean photon number $n$. The covariance matrix measured after amplification is given by $\tilde{V}$, while $V$ represents the quantum state. Thus a good estimate of $\sqrt{G}$ and $n$ would allow us to reconstruct the quantum statistics.

\subsection{Two-mode squeezed state} \label{appendix: tms estimate}
There are different methods to calibrate gain and added noise, which typically require some form of calibrated noise source \cite{Chang2018} or a temperature sweep \cite{Tholen2009, Mariantoni2010}. A method to roughly estimate the gain of the amplification chain using only our device is by measuring cross-correlations in a two-mode squeezed state. Assuming the added noise is thermal, the cross-correlations corresponding to squeezing are independent of the added noise but not the amplifier gain. 
We quantify these correlations by adding the relevant elements of the covariance matrix as
\begin{equation}
    C = \sqrt{\tilde{V}^2_{13} + \tilde{V}^2_{14} +\tilde{V}^2_{23} + \tilde{V}^2_{24} }. \label{eq: C_def}
\end{equation}

\begin{figure}[htbp]
	\centering
	\includegraphics[scale=0.55]{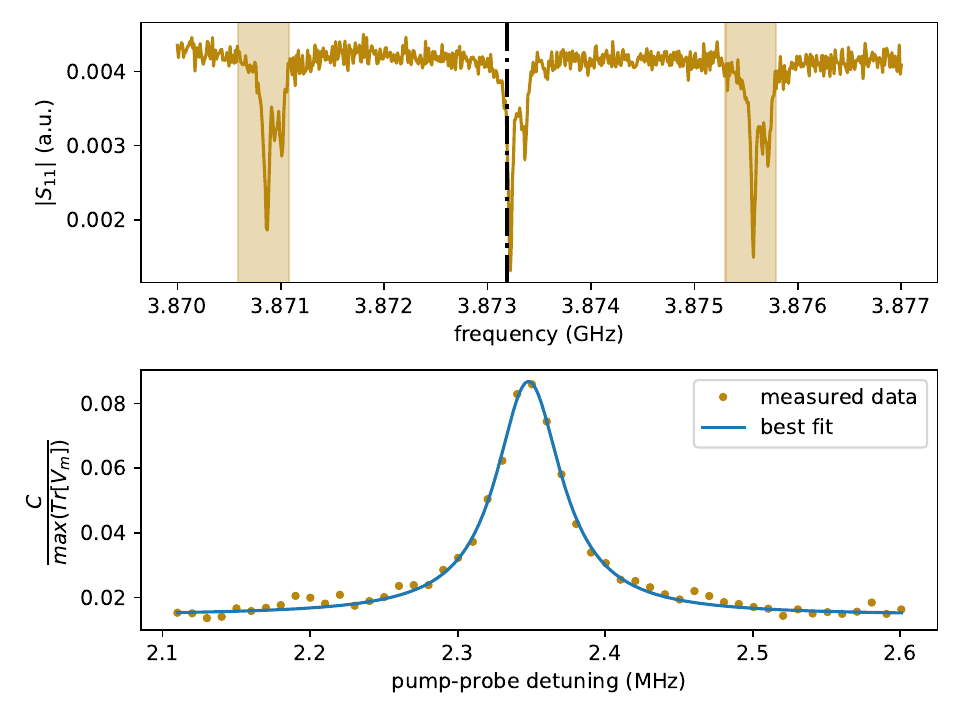}
	\caption{\textbf{Gain estimation.} Top: Frequency sweep depicting the three SAW modes of interest. A pump is positioned at the vertical black dashed line, while noise is measured at frequencies symmetric around the pump, as indicated by the shaded region.
	Bottom: The correlation quantity $C$ is plotted as a function of pump-probe detuning. A fit to $C$ is made to extract the gain of our amplification chain, assuming an effective phonon temperature of \SI{30}{mK}. From the fit we obtain $G \approx \SI{80}{dB}$. The fit also provides $|\epsilon| \approx \SI{6}{kHz}$. For comparison, $C$ is scaled by the maximum trace $\textrm{max}(\textrm{Tr}[\tilde{V}])$ measured across all frequencies.}
	\label{fig: corrsweep}
\end{figure}

We measure $C$ by applying a flux pump at 3.8732 GHz, while measuring the noise in a pair of neighbouring modes. As illustrated in Fig. \ref{fig: corrsweep}, the pump frequency is placed directly on a SAW mode and probe frequencies are swept across neighbouring SAW modes, always keeping them strictly symmetric with respect to the pump to satisfy the 4-wave mixing criterion. In Fig. \ref{fig: corrsweep}, $C$ is plotted as a function of the detuning between the pump and the probe frequencies. The probe frequency sweep results in a Lorentzian-like shape of $C$ as a function of detuning, where the peak occurs when both probes are located within their respective SAW modes.

\begin{figure}[h]
    \centering

\begin{tikzpicture}[->,>=stealth',shorten >=1pt,auto,node distance=2.2cm,
                    semithick]
  \tikzstyle{every state}=[fill=gray,draw=none,text=white]

  \node[state]         (A)              {$a_1$};
  \node[state]         (B) [right of=A]  {$a_2^\dagger$};

  \path (B) edge [bend left] node {$-\epsilon$} (A);
  \path (A) edge [bend left] node {$\epsilon^*$} (B);

  \path (B) edge [loop right] node {$-\Delta_2^*$} (B);
  \path (A) edge [loop left] node {$\Delta_1$} (A);
  
\end{tikzpicture}

    \caption{\textbf{Graph for two-mode squeezing}. The graph provides us with the mode-coupling matrix $M$. The covariance matrix $V^{\text{out}}$ can then be calculated according to Eq. \eqref{eq: V(S)} or Eq. \eqref{eq: V(S) w losses}. }
    \label{fig: two_mode_graph}
\end{figure}
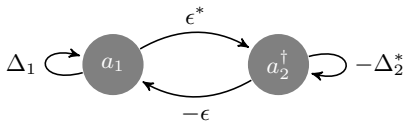

The $C$ lineshape is calculated by deriving the covariance matrix $\tilde{V}$ for two coupled SAW modes, according to the method outlined in Appendix \ref{appendix: theory cov matrix}.
More specifically, the mode-coupling matrix $M$ for two modes is graphically represented in Fig \ref{fig: two_mode_graph}. The outgoing noise is then fully characterized by the $4 \times 4$-matrix $V^{\text{out}}$, which we can find by following Eq. \eqref{eq: scattering matrix} - \eqref{eq: V(S) w losses}. Amplification is taken into account by substituting  $V \rightarrow V^{\text{out}}$ in Eq. \eqref{eq: amp_transform}. The resulting covariance matrix $\tilde{V}$ is used to calculate $C$.

Given the resonance frequencies of the SAW modes along with their linewidths and assuming $\tilde{g}_1 = \tilde{g}_2$, we are left with three unknown parameters: the gain $G$, parametric coupling $\epsilon$ and the effective phonon temperature $T_\text{eff}$. If we fix the phonon temperature, a fit to $C$ will give us the gain $G$ and the parametric coupling $\epsilon$.
An example fit is shown as a solid line in Fig.~\ref{fig: corrsweep}, with an assumed phonon temperature of $T_\text{eff} = \SI{30}{mK}$. This yields an estimate of the gain to be $G\approx \SI{80}{dB}$, which lies within range of our expectations.

This method is not a substitute for proper gain and noise calibration procedures. However, we will use this method to make an estimate on the maximal phonon temperature possible for the two-mode squeezing in Fig. \ref{fig: twomode_squeezing} to be a signature of entanglement.
The procedure consists of essentially three steps: extract $G$ by fitting to $C$, estimate the added noise $N$ and finally reconstruct the original quantum statistics from data in Fig. \ref{fig: twomode_squeezing} according to Eq. \eqref{eq: amp_transform}.

After the gain $G$ is extracted from fitting to $C$, the amplifier noise is estimated by solving for $n$ in Eq.~\eqref{eq: amp_transform}, by replacing $\tilde{V}$ by the pump off statistics $V_\text{off}$ while $V$ is substituted by a thermal state with the corresponding temperature $T_\text{eff}$.
Finally, solving for $n$ with these assumptions yields $n \approx 0.08$.
This added noise value is a very low estimate.
Using a higher added noise value during reconstruction of $V$ however, will result in more squeezing.

With the gain $G$ and noise $n$, one can attempt reconstructing the pre-amplified covariance matrix from the two-mode squeezing data in Fig. \ref{fig: twomode_squeezing}.
To determine whether the reconstructed state is entangled, we apply the partial positive transpose (PPT) criterion \cite{Weedbrook2012, Simon2000}.
If the reconstructed two-mode squeezed state is labelled $V_\text{TMS}$, the PPT criteria states that if we do a partial transposition:
\begin{align}
    \Lambda &= \text{diag}(1, 1, 1, -1),
    \\
    \bar{V}_\text{TMS} &= \Lambda V_\text{TMS} \Lambda,
\end{align}
then a necessary and sufficient condition for separability for bipartite Gaussian states is that the matrix $H = \bar{V}_\text{TMS} + i \Omega$ is positive semidefinite.
$\Omega$ is the symplectic matrix, defined as
\begin{align}
    \Omega &= \begin{pmatrix}
        0 & 1 & 0 & 0 \\
        -1 & 0 & 0 & 0 \\
        0 & 0 & 0 & 1 \\
        0 & 0 & -1 & 0
        \end{pmatrix}.
        \label{eq: symplecticdef}
\end{align}
Thus we test for entanglement by calculating the eigenvalues $\lambda$ of $H$ and checking whether the smallest eigenvalue $\lambda_\text{min}$ is negative.
Note that for $\Omega$ and $\Lambda$ we are assuming the covariance matrix is in the basis $(I_1, Q_1, I_2, Q_2)$.

\begin{figure}[htbp]
	\includegraphics[scale=0.55]{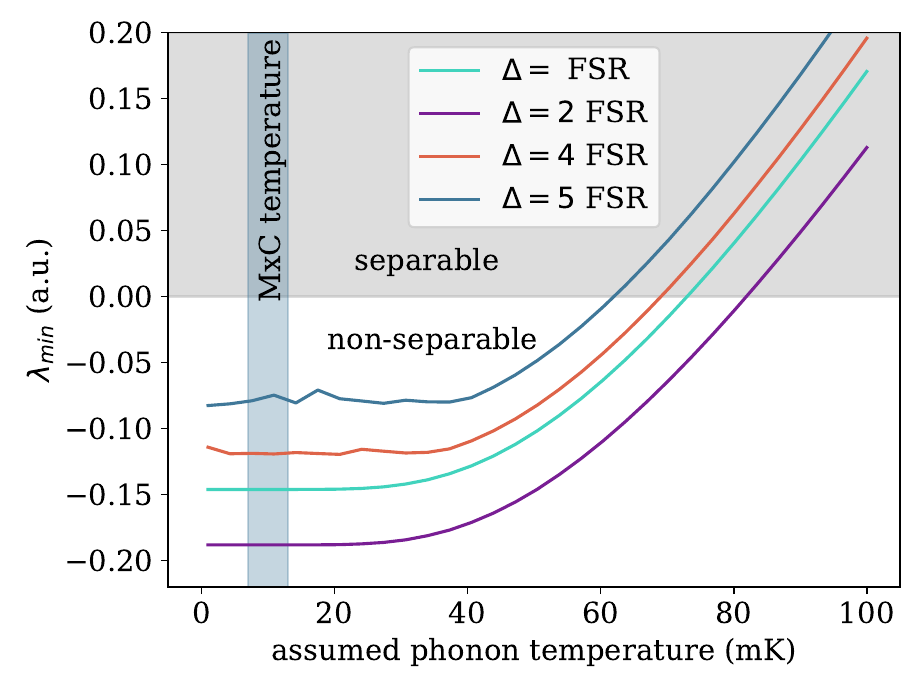}
	\caption{\textbf{PPT criterion results.} We apply the PPT criterion to data presented in Fig. \ref{fig: twomode_squeezing} in the main text. 
	We vary the assumed phonon temperature, estimate $G$ and $n$, and evaluate the PPT criteria at each step.
	The results suggest entanglement is present up to a phonon temperature of roughly \SI{63}{mK}. Each line corresponds to different detunings from the pump.}
	\label{fig: ppt_test}
\end{figure}

However, the eigenvalue $\lambda_\text{min}$ depends on the phonon temperature, since it affects our estimate of $G$ and $n$.
We take this into account by calculating the value of $\lambda_{min}$ at various phonon temperatures $T_\text{eff}$, presented in Fig.~\ref{fig: ppt_test}.
Accordingly, we observe that entanglement persists for phonon temperatures up to \SI{63}{mK}.
This should be compared to the mixing chamber temperature of roughly \SI{10}{mK} and previous experiments estimating the effective SAW phonon temperature to \SI{37}{mK} \cite{Noguchi2017}.
Together, these observations suggest that the measured two-mode squeezing is a signature of entangled SAW modes.

\subsection{Calibration and multimode state reconstruction}
\label{appendix: calibration}
\begin{figure}[htbp]
	\centering
	\includegraphics[scale=0.66]{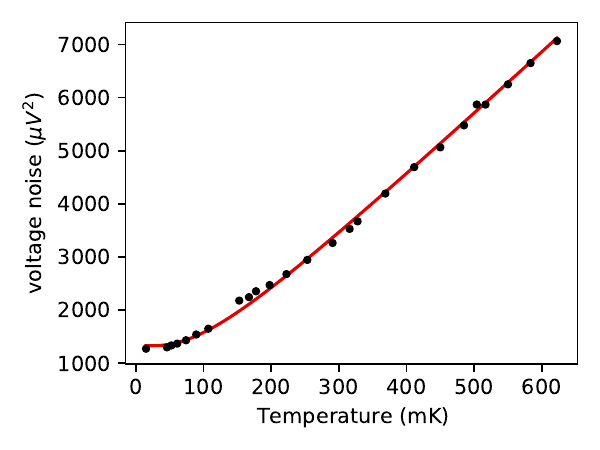}
	\caption{\textbf{Planck spectroscopy calibration.} Noise power as a function of temperature at the frequency of mode $f_2$ (\SI{3.858}{GHz}).}
	\label{fig: calnoise}
\end{figure}

For the multimode entanglement experiment we perform a calibration of the gain and added noise in the amplification chain. We substitute a resistor at the mixing chamber for the device and measure the noise power as a function of temperature. Fits to the expression 
\begin{equation}
        P = G h f  \left[ \frac{1}{2}\coth{\frac{h f}{2 k_B T_\mathrm{mxc}}} + \frac{1}{2}(2n +1) \right]
\end{equation}
give the gain and noise parameters. The calibration is performed without the resonator connected at each frequency used in the measurement and the heating raises the temperature of the entire mixing chamber stage of the cryostat. Figure~\ref{fig: calnoise} shows the noise power at the frequency of mode $f_2$ (cf. Fig.~\ref{fig:  twopump covar}). As the amplification chain makes use of a parametric amplifier, care must be taken to account for the idler noise in the analysis \cite{Malnou2021, ranadive2021}.

To accurately account for the variation in gain with frequency, the gain and noise terms appearing in Eq.~\ref{eq: amp_transform} are extended to $T = \bigoplus_{i=1}^n \sqrt{G_i} I$ and $N = \bigoplus_{i=1}^n (G_i-1)(2 n+1)I + (G_{I,i}-1)(2 n_{I,i}+1)I$. The index $i$ denotes the frequency modes and the subscript $I$ denotes the idler contribution. To obtain reasonable fit parameters, we restrict the signal-idler gain to $G_{I,i}=G_i$ and assume an idler noise to originate from a thermal state at temperature $T_I = \SI{30}{mK}$. From our analysis we obtain a lower than expected added noise temperature of $T_n \leq \SI{300}{mK}$. In case our calibration underestimates the real added noise in the amplification chain, that should imply \emph{more} entanglement in the reconstructed multimode state as the added noise obscures the correlations.

The fits yield uncertainties for the estimated gain and noise, which influence the entanglement significance by error propagation. The error accounting for the calibration as well as measurement error can be written as
\begin{equation}
    \sigma^2_{ij} = \sigma^2_{ij,A} +\sigma^2_{ij,B} + \sigma^2_{ij,C}.
\end{equation}
The contribution related to uncertainty in the signal gain is given by
\begin{equation}
\begin{split}
    \sigma^2_{ij,A} &= \left[\left(\frac{\tilde{V}_{ij}}{2\sqrt{G_{i}^3G_{j}}}\sigma_{G_{i}}\right)^2 + \left(\frac{\tilde{V}_{ij}}{2\sqrt{G_{j}^3G_{i}}} \sigma_{G_{j}}\right)^2 \right] \left(1+\delta_{ij}\right) \\ 
    &+ 2\delta_{ij}\left(\frac{2n_i+1}{G_{i}}\sigma_{G_{i}}\right)^2.
\end{split}
\end{equation}
Additional contributions arise from uncertainty in the added noise $\sigma_{ii,B}= 2\sigma_{n_i}$ as well as the measured fluctuations in the covariance matrix
\begin{equation}
    \sigma_{ij,C} = \frac{\sigma_{\tilde{V}_{ij}}}{\sqrt{G_i G_j}}.
\end{equation}
Here, $\sigma_{\tilde{V}_{ij}}$ is given by the standard error of the mean of each covariance matrix element as calculated from the measured data. Furthermore, we  cannot assume that errors in gain and noise are uncorrelated, which leads to the diagonal error term
\begin{equation}
    \sigma^2_{ij,\mathrm{corr}} = 4\frac{V_{ii}-(2n_i+1)}{G_i}\mathrm{Cov}\left(G_i,n_i\right) \delta_{ij}.
\end{equation}

The covariance matrix for a physical state must satisfy the Heisenberg uncertainty relations, which may be expressed as
\begin{align}
    V \geq & \ 0, \label{eq: constaint1} \\
    V - i\Omega \geq & \ 0. \label{eq: constaint2}
\end{align}
where $\Omega$ is the symplectic matrix (cf. Eq.~\ref{eq: symplecticdef}).
Due to measurement noise and drift, this is not guaranteed to hold for the covariance matrix $V$ obtained by inverting Eq.~\ref{eq: amp_transform}. To ensure a physical state before applying entanglement tests, we apply a reconstruction to find the most probable physical state $V$ given a noisy measured state $V'$ \cite{Shchukin_2016}. This $V$ is obtained by solving the optimization problem
\begin{align}
    \min_V \left( \max_{\alpha\beta}\frac{ \left|V'_{\alpha\beta} - V_{\alpha\beta} \right|}{\sigma_{\alpha\beta}} \right).
    \label{eq: reconstruction}
\end{align}
The pump configuration used in our measurement would suggest including more probe modes at higher frequency in the analysis. Including these modes renders the covariance matrix obtained unphysical by multiple standard deviations, presumably due to error in our calibration.

\section{Two-mode quadrature histograms}
\begin{figure}[!htb]
	\centering
	\includegraphics[scale=0.66]{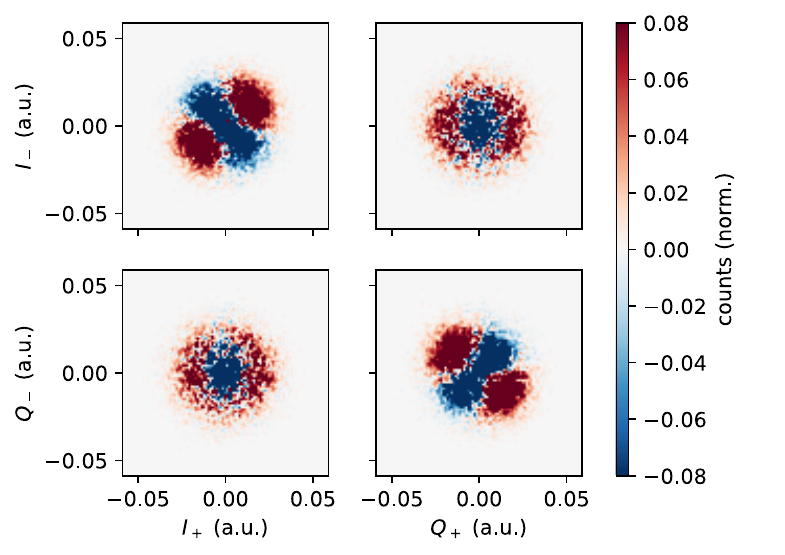}
	\caption{{\textbf{Single pump two-mode quadrature histograms.} Measured for the mode pair adjacent to the pump mode ($\Delta=FSR$). The pump off data has been subtracted. The squeezing appears in the planes combining the $I$ or $Q$ quadratures of both modes. Plotting the $I$ and $Q$ quadratures of different modes shows amplified noise, resulting in donut-shaped subtracted histograms.}}
	\label{fig: histograms_singlepump}
\end{figure}

{Figure~\ref{fig: twomode_squeezing}b shows the two-mode quadrature histograms in the $I_{+}-I_-$ plane. Here, the squeezing axis corresponds to the diagonal. In Fig.~\ref{fig: histograms_singlepump} we plot all two-mode quadrature histograms for the mode pair closest to the pump. The squeezing is manifest also in the $Q_{+}-Q_-$ histogram, while the other quadratures show amplified noise.}

\begin{figure}[!htb]
	\centering
	\includegraphics[scale=0.66]{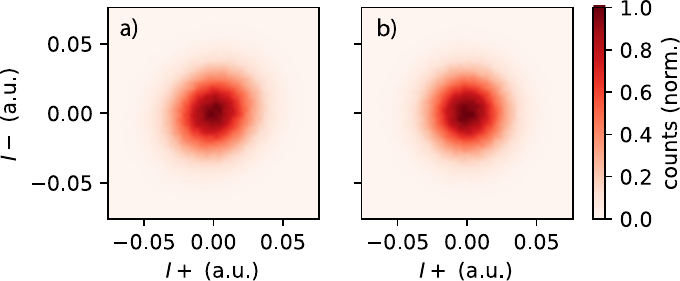}
	\caption{\textbf{Two-mode quadrature histograms without subtraction.} The histogram obtained with the pump on is shown in \textbf{a}, while \textbf{b} shows the reference histogram measured with the pump off.}
	\label{fig: histograms_singlepump_nosub}
\end{figure}

The $I_{+}-I_-$ histogram is shown without subtraction in Fig.~\ref{fig: histograms_singlepump_nosub}a. In Fig.~\ref{fig: histograms_singlepump_nosub}b we plot the reference histogram obtained with the pump off that is subtracted to generate the data shown in Fig.~\ref{fig: histograms_singlepump}.

\FloatBarrier
\hspace{1cm}
\section{Scattering measurements}
\label{appendix: scattering}

\begin{figure}[htbp]
	\centering
	\includegraphics[scale=0.66]{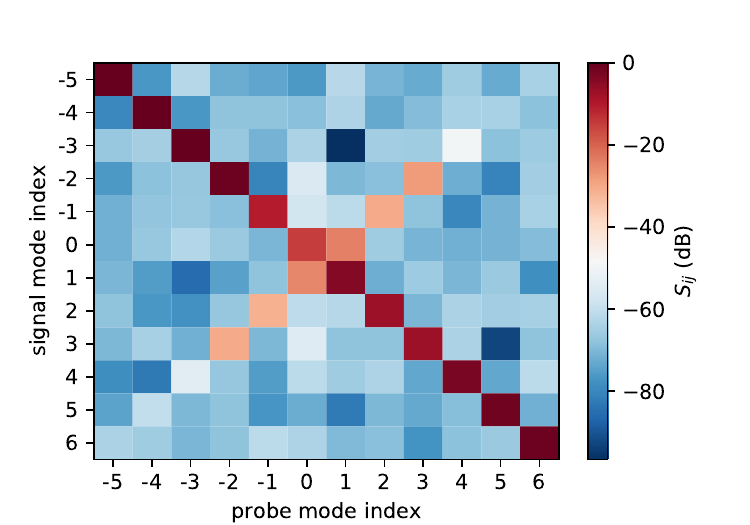}
	\caption{\textbf{Scattering matrix.} Magnitude of scattering matrix elements measured with a single pump tone. Scattering occurs between modes symmetric in frequency around the pump. The mode indices are consistent with Fig.~\ref{fig: twopump covar}. The dB scale is referenced to the reflected amplitude in mode $-5$. }
	\label{fig: scatteringmatrix_singlepump}
\end{figure}

\begin{figure}[htbp]
	\centering
	\includegraphics[scale=0.66]{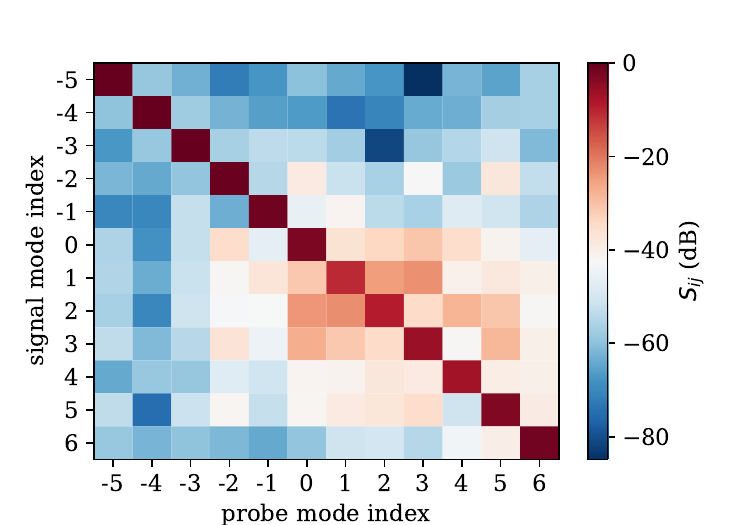}
	\caption{\textbf{Scattering matrix.} Magnitude of scattering matrix elements measured with four pump tones. The mode indices are consistent with Fig.~\ref{fig: twopump covar}. The dB scale is referenced to the reflected amplitude in mode $-5$. }

	\label{fig: scatteringmatrix}
\end{figure}

To verify the mode couplings induced by parametric pumping we perform scattering measurements in sample B. A signal is injected into one mode via the IDT and the scattering into other modes is measured. The scattering matrix for a single pump tone (four tones) is shown in Fig.~\ref{fig: scatteringmatrix_singlepump} (Fig.~\ref{fig: scatteringmatrix}). The scattering measurements verify that the parametric couplings relied on to generate entanglement are present. For an evenly-spaced pump comb the couplings are not all-to-all due to the deviation from uniform SAW mode spacing.

\begin{figure}[htbp]
	\centering
	\includegraphics[scale=0.66]{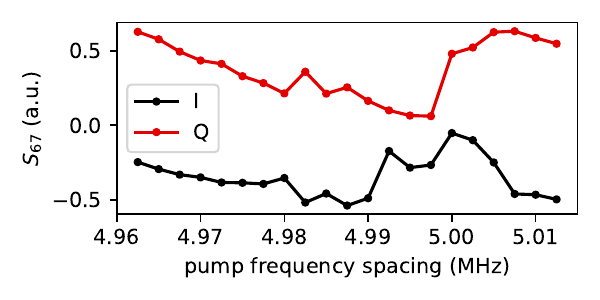}
	\caption{\textbf{Scattering variation with pump frequency.} The $I$ and $Q$ quadrature amplitudes of the $S_{67}$ scattering matrix element as a function of pump tone spacing. The four-tone pump comb is uniform in frequency spacing and amplitude.}
	\label{fig: S67}
\end{figure}

The phase and amplitude of the scattered signal is sensitive to the pump configuration. Figure~\ref{fig: S67} shows the $I$ and $Q$ quadratures of a single scattering matrix element as a function of the pump spacing. 

\end{document}